\def\year{2018}\relax
\documentclass[letterpaper]{article} 
\usepackage{aaai18}  
\usepackage{times}  
\usepackage{helvet}  
\usepackage{courier}  
\usepackage{url}  
\usepackage{graphicx}  
\usepackage{booktabs} 
\usepackage{subfig}

\newcommand{\citet}[1]{\citeauthor{#1}~\shortcite{#1}}

\begin{document}
\title{Beyond Views: Measuring and Predicting Engagement in Online Videos}
\author{Siqi Wu \and Marian-Andrei Rizoiu \and Lexing Xie\\
Australian National University and Data 61, CSIRO, Australia\\
{\{siqi.wu, marian-andrei.rizoiu, lexing.xie\}}@anu.edu.au\\
}

\maketitle

\begin{abstract}
The share of videos in the internet traffic has been growing, therefore understanding how videos capture attention on a global scale is also of growing importance. 
Most current research focus on modeling the number of views, but we argue that video engagement, or time spent watching is a more appropriate measure for resource allocation problems in attention, networking, and promotion activities. 
In this paper, we present a first large-scale measurement of video-level aggregate engagement from publicly available data streams, on a collection of 5.3 million YouTube videos published over two months in 2016.
We study a set of metrics including time and the average percentage of a video watched.
We define a new metric, \textit{relative engagement}, that is calibrated against video properties and strongly correlate with recognized notions of quality.
Moreover, we find that engagement measures of a video are stable over time, thus separating the concerns for modeling engagement and those for popularity -- the latter is known to be unstable over time and driven by external promotions. 
We also find engagement metrics predictable from a \textit{cold-start} setup, having most of its variance explained by video context, topics and channel information -- $R^2$=0.77. 
Our observations imply several prospective uses of engagement metrics -- choosing engaging topics for video production, or promoting engaging videos in recommender systems.
\end{abstract}

\section{Introduction}
\label{sec:intro}

Attention is a scarce resource in the modern world. 
There are many metrics for measuring attention received by online content, such as page views for webpages, listen counts for songs, view counts for videos, and the number of impressions for advertisements.
Although these metrics describe the human behavior of \textit{choosing} one particular item, they do not describe how users \textit{engage} with this item~\cite{van2016aligning}.
For instance, an audience may become immersed in the interaction or quickly abandon it -- the distinction of which will be clear if we know how much time the user spent interacting with this given item.
Hence, we consider popularity and engagement as different measures of online behavior.

In this work, we study online videos using publicly available data from the largest video hosting site YouTube.
On YouTube, popularity is characterized as the willingness to click a video, whereas engagement is the watch pattern after clicking.
While most research have focused on measuring popularity~\cite{pinto2013using,rizoiu2017expecting}, engagement of online videos is not well understood, leading to key questions such as:
How to measure video engagement?
Does engagement relate to popularity?
Can engagement be predicted?
Once understood, engagement metrics will become relevant targets for recommender systems to rank the most valuable videos.

\begin{figure}[t]
    \centering
    \includegraphics[width=0.47\textwidth]{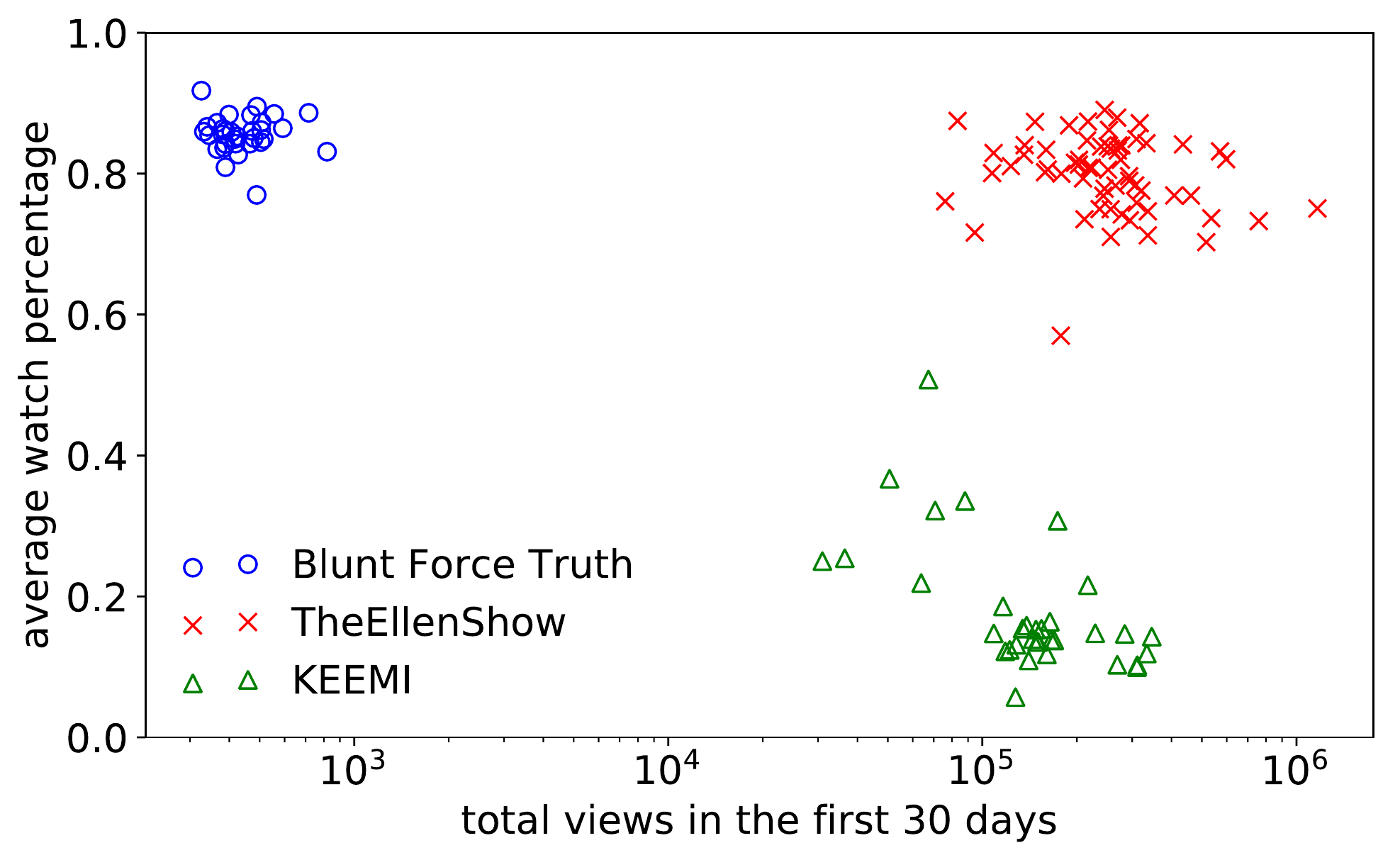}
    \caption{Scatter plot of videos from three YouTube channels: \texttt{Blunt Force Truth} (political entertainment, blue circle), \texttt{KEEMI} (cooking vlog, green triangle), and \texttt{TheEllenShow} (comedy, red cross). x-axis: total views in the first 30 days; y-axis: average watch percentage.}
    \label{fig:teaser}
\end{figure}

In Fig.~\ref{fig:teaser}, we plot the number of views against the average percentage watched for 128 videos in 3 channels.
While the entertainment channel \texttt{Blunt Force Truth} has the least views on average, the audience tend to watch more than 80\% of each video.
On the contrary, videos from the cooking vlogger \texttt{KEEMI} have on average 159,508 views, but they are watched only 18\%.
This example illustrates that videos with a high number of views do not necessarily have high watch percentages, and prompts us to investigate other metrics for describing engagement.

Recent progress in understanding video popularity and the availability of new datasets allow us to address three open questions about video engagement.
Firstly, \textbf{on an aggregate level, how to measure engagement?}
Most engagement literatures focus on the perspective of an individual user, such as recommending relevant products~\cite{covington2016deep}, tracking mouse gestures~\cite{arapakis2014understanding} or optimizing search results~\cite{drutsa2015future}.
Since user-level data is often unavailable, defining and measuring average engagement is useful for content producers on YouTube.
Secondly, \textbf{within the scope of online video, can engagement help measure content quality?}
As shown in Fig.~\ref{fig:teaser}, video popularity metric is inadequate to estimate quality.
One early attempt to measure online content quality was taken by~\citet{salganik2006experimental}, who studied music listening behavior in an experimental environment.
For a large number of online contents, measuring quality from empirical data still remains unexplored.
Lastly, \textbf{in a \textit{cold-start} setup, can engagement be predicted?}
Online attention is known to be difficult to predict without early feedback~\cite{martin2016exploring}.
For engagement, \citet{park2016data} showed the predictive power of user reactions such as views and comments.
However, these features also require monitoring the system for a period of time.
In contrast, if engagement can be predicted before content is uploaded, it will provide actionable insights to content producers.

We address the first question by constructing 4 new datasets that contain more than 5 million YouTube videos.
We build two 2-dimensional maps that visualize the internal bias of existing engagement metrics -- \textit{average watch time} and \textit{average watch percentage} -- against video length.
Building upon that, we derive a novel \textit{relative engagement} metric, as the duration-calibrated rank of average watch percentage.

Answering the second question, we demonstrate that relative engagement is stable over time, and strongly correlates with established quality measures in Music and News categories, such as Billboard songs, Vevo artists, and top news channels.
This newly proposed relative engagement metric can be a target for recommender systems to prioritize quality videos, and for content producers to create engaging videos.

Addressing the third question, we predict engagement metrics in a \textit{cold-start} setting, using only video content and channel features. 
With off-the-shelf machine learning algorithms, we achieve $R^2$=0.77 for predicting average watch percentage.
We consider this as a significant result that shows the predictability of engagement metrics.
Furthermore, we explore the predictive power of video topics and find some topics are strong indicators for engagement. 

The main contributions of this work include: 
\begin{itemize}
\item Conduct a large-scale measurement study of engagement on 5.3 million videos over two-month period, and publicly release 4 new datasets and the engagement benchmarks\footnote{The code and datasets are publicly available at \url{https://github.com/avalanchesiqi/youtube-engagement}}.
\item Measure a set of engagement metrics for online videos, including average watch time, average watch percentage, and a novel metric -- relative engagement, which is calibrated with respect to video length, stable over time, and correlated with video quality. 
\item Predict relative engagement and watch percentage from video context, topics, and channel reputation in a \textit{cold-start} setting (i.e., before the video gathers any view or comment), achieving $R^2$=0.45 and 0.77 respectively. 
\end{itemize}

\section{Datasets}
\label{sec:data}

We curate 4 new publicly available video datasets, as summarized in Table~\ref{table:data} and Table~\ref{table:category}.
We also describe three daily series available for all videos: shares, views and watch time.

\subsection{Video datasets}
\label{ssec:datasets}

\begin{table}
    \centering
    \begin{tabular}{rrr}
        \toprule
        Dataset & \#Videos & \#Channels\\
        \midrule
        \textsc{Tweeted Videos} & 5,331,204 & 1,257,412 \\
        \textsc{Vevo Videos} & 67,649 & 8,685\\
        \textsc{Billboard Videos} & 63 & 47\\
        \textsc{Top News Videos} & 28,685 & 91\\
        \bottomrule
    \end{tabular}
    \caption{Overview of 4 new video datasets.}
    \label{table:data}
\end{table}

\begin{table}
    \centering
    \begin{tabular}{rrrr}
        \toprule
        Category & \#Videos & Category & \#Videos  \\
        \midrule
        People & 1,265,805 & Comedy & 138,068 \\
        Gaming & 1,079,434 & Science & 110,635 \\
        Entertainment & 775,941 & Auto & 84,796 \\
        News & 459,728 & Travel & 65,155 \\
        Music & 449,314 & Activism & 58,787\\
        Sports & 243,650 & Pets & 27,505 \\
        Film & 194,891 & Show & 1,457 \\
        Howto & 192,931 & Movie & 158 \\
        Education & 182,849 & Trailer & 100 \\
        \bottomrule
    \end{tabular}
    \caption{Breakdown of \textsc{Tweeted Videos} by category.}
    \label{table:category}
\end{table}

\noindent\textbf{\textsc{Tweeted videos}} contains 5,331,204 videos published between July 1st and August 31st, 2016 from 1,257,412 channels.
The notion of \textit{channel} on YouTube is analogous to that of \textit{user} on other social platforms, since every video is published by a channel and belonging to one user account.
Using Twitter mentions to sample a collection of YouTube videos has been used in previous works~\cite{abisheva2014watches,yu2014twitter}.
We use the Twitter Streaming API to collect tweets, by tracking the expression \textsc{"youtube" OR ("youtu" AND "be")}.
This covers textual mentions of YouTube, YouTube links and YouTube's URL shortener (\url{youtu.be}).
This yields 244 million tweets over the two-month period.
In each tweet, we search the \texttt{extended\_urls} field and extract the associated YouTube video ID.
This results in 36 million unique video IDs and over 206 million tweets.
For each video, we extract its metadata and three attention-related dynamics, as described in Sec.~\ref{ssec:attention-dynamics}.
A non-trivial fraction (45.82$\%$) of all videos have either been deleted or their statistics are not publicly available.
This leaves a total of 19.5 million usable videos.

We further filter videos based on recency and their level of attention.
We remove videos that are published prior to this two-month period to avoid older videos, since being tweeted a while after being uploaded may indicate higher engagement.
We also filter out videos that receive less than 100 views within their first 30 days after upload, which is the same filter  used by~\citet{Brodersen2012a}.
Videos that do not appear on Twitter, or have extremely low number of early views are unlikely to accumulate a large amount of attention~\cite{rizoiu2017online,pinto2013using}, therefore, they do not provide enough data to reflect collective watch patterns.
Our proposed measures can still be computed on these removed videos, however the results might have limited relevance given the low level of user interaction with them.
Table~\ref{table:category} shows a detailed category breakdown of \textsc{Tweeted videos}.

\noindent\textbf{\textsc{Quality videos}.} We collect three datasets containing videos deemed of high quality by domain experts, two of which are on Music and one is on News.
These datasets are used to link engagement and video quality (Sec~\ref{ssec:video-quality}).

\textbf{$\bullet$ \textsc{Vevo Videos}.}
Vevo is a multinational video hosting service which syndicates licensed music clips from three major record companies on YouTube~\cite{vevowikipedia}.
\textsc{Vevo} artists usually come from professional music background, and their videos are professionally produced.
We consider \textsc{Vevo Videos} to be of higher quality than the average Music videos in the \textsc{Tweeted Videos} dataset.
We collect all the YouTube channels that contain the keyword ``Vevo'' in the title and a ``verified'' status badge on the profile webpage.
In total, this dataset contains 8,685 Vevo channels with 67,649 music clips, as of August 31st, 2016.

\textbf{$\bullet$ \textsc{Billboard Videos}.}
Billboard acts as a canonical ranking source in the music industry, aggregating music sales, radio airtime and other popularity metrics into a yearly Hot 100 music chart.
The songs that appear in this chart are usually perceived as having vast success and being of high quality.
We collect 63 videos from 47 artists based on the 2016 Billboard Hot 100 chart~\cite{billboardwikipedia}.

\textbf{$\bullet$ \textsc{Top News Videos}} features a list of top 100 most viewed News channels, as reported by an external ranking source~\cite{top100news}.
This list includes traditional news broadcasting companies (e.g., \texttt{CNN}), as well as popular online talk shows (e.g., \texttt{The Young Turks}).
For each channel, we retrieve its last 500 videos published before Aug 31st, 2016. 
This dataset contains 91 publicly available News channels and 28,685 videos.

\subsection{YouTube metadata and attention dynamics}
\label{ssec:attention-dynamics}

For each video, we use the YouTube Data API to retrieve video metadata information -- video id, title, description, upload time, category, duration, definition, channel id, channel title and associated Freebase topic ids, which we resolve to entity names using the latest Freebase data dump\footnote{\url{https://developers.google.com/freebase}}~\cite{figueiredo2014does}.

We then develop a software package\footnote{\url{https://github.com/computationalmedia/youtube-insight}} to extract three daily series of video attention dynamics: daily volume of shares, view counts and watch time.
Throughout this paper, we denote the number of shares and views that a video receives on the $t^{th}$ day after upload as $s[t]$ and $x_v[t]$, respectively.
Similarly, $x_w[t]$ is the total amount of time of video being watched on the $t^{th}$ day.
Each attention series is observed for at least 30 days, i.e., $t = 1, 2, \ldots 30$.
Most prior research on modeling video popularity dynamics~\cite{szabo2010predicting,figueiredo2016trendlearner} study only view counts. 
To the best of our knowledge, our work is the first to perform large-scale measurements on video watch time.

\section{Measures of video engagement}
\label{sec:measure}

In this section, we measure the interplay between view count, watch time, watch percentage and video duration.
We first examine their relation in a new visual presentation -- \textit{engagement map}, then we propose \textit{relative engagement}, a novel metric to estimate video engagement (Sec.~\ref{ssec:engagement-score}).
We show that relative engagement calibrates watch patterns for videos of different lengths, demonstrates correlation to external notions of video quality (Sec.~\ref{ssec:video-quality}), and remains stable over time (Sec.~\ref{ssec:temp-dynamics}).

\subsection{Discrepancy between views and watch time}
\label{ssec:comp-view-watch}

\begin{figure}[t]
    \centering
    \includegraphics[width=0.47\textwidth]{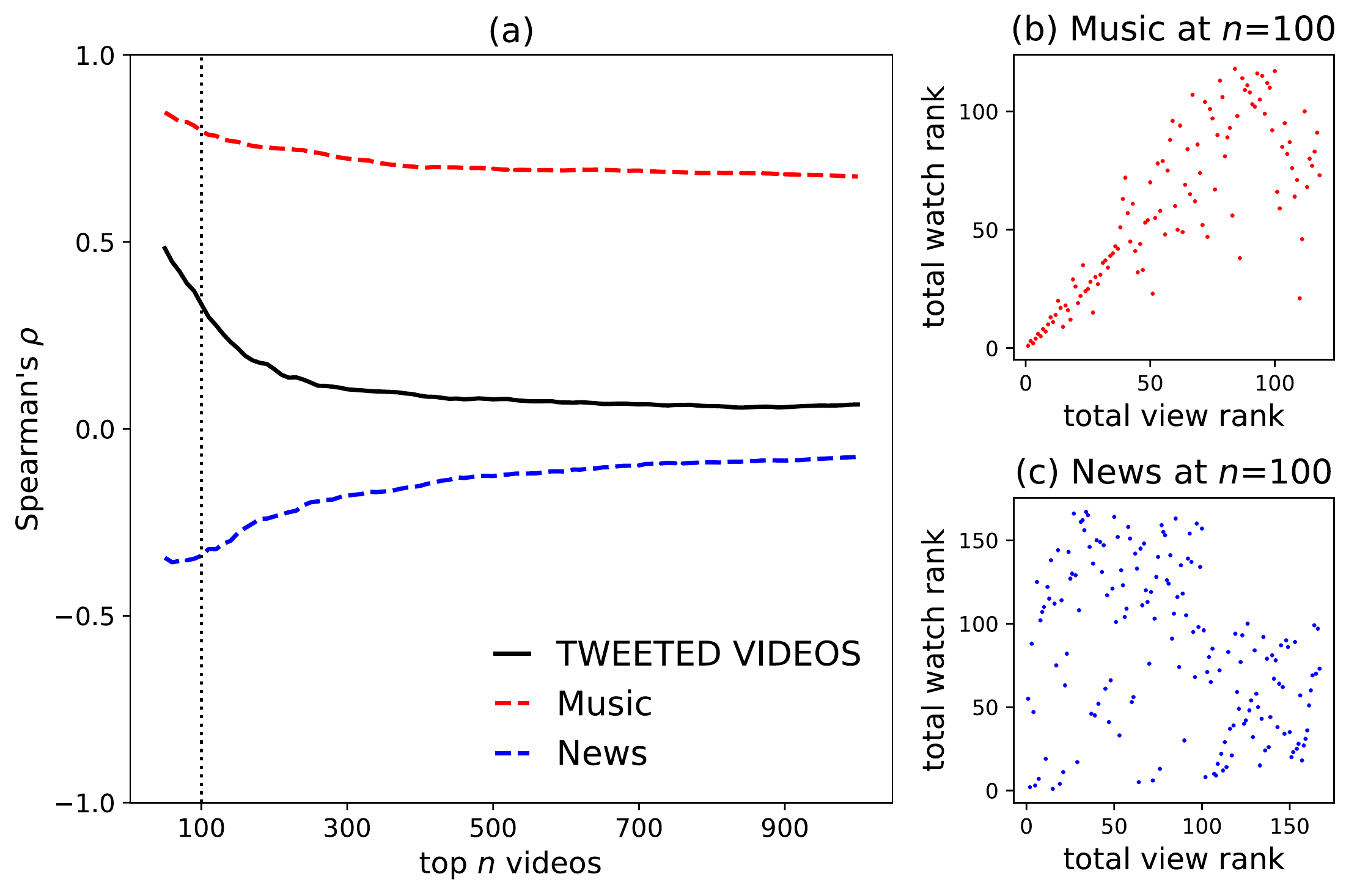}
    \caption{(a) Disagreement between the union set of top $n$ most viewed and top $n$ most watched videos in \textsc{Tweeted videos} at the age of 30 days, measured with Spearman's $\rho$. (b-c) Scatter plots of video ranking in view and in watch at $n$=100 in Music ($\rho$=0.80) and News ($\rho$=-0.34).}
    \label{fig:view-watch-diff}
\end{figure}

Fig.~\ref{fig:teaser} illustrates that watch patterns (e.g., average percentage of video watched) can be very different for videos with similar views.
We examine the union set of top $n$ videos in \textsc{Tweeted Videos} dataset, respectively ranked by total views and total watch time at the age of 30 days.
For $n$ varying from 100 to 1000, we measure their agreement using Spearman's $\rho$.
With value between -1 and +1, a positive $\rho$ implies that as the rank in one variable increases, so does the rank in the other variable.
A $\rho$ of 0 indicates no correlation exists in these two ranked variables.
Fig.~\ref{fig:view-watch-diff}a shows that in \textsc{Tweeted videos}, video ranks in total view count and total watch time correlate at the level of 0.48 when $n$ is 50, but this correlation declines to 0.08 when $n$ increases to 500 (solid black line).
Furthermore, the level of agreement varies across different video categories: for Music, a video that ranks high in total view count often ranks high in total watch time ($\rho$=0.80 at $n$=100, Fig.~\ref{fig:view-watch-diff}b);
for News, the two metrics have a weak negative correlation ($\rho=-0.34$ at $n=100$, Fig.~\ref{fig:view-watch-diff}c).

This observation indicates that total view count and total watch time provide different aspects of how audience interact with YouTube videos.
One recommender system optimizing for view count may generate remarkably different results with one that drives watch time~\cite{yi2014beyond}.
In the next section, we analyze their interplay to construct more diverse set of measures for video engagement.

\subsection{Engagement map and relative engagement}
\label{ssec:engagement-score}

Recent studies show that the quality of a digital item is linked to the audience's decision to continue watching or listening after first opening it ~\cite{salganik2006experimental,krumme2012quantifying}.
Therefore, the average amount of time that the audience spend on watching a video should be indicative of video quality.
For a given video, we compute two aggregate metrics:

\begin{itemize}
    \item \textit{average watch time}$\bar\omega_t$: the total watch time $x_w[1: t]$ divided by the total view count $x_v[1: t]$ up to day $t$ 
    \begin{equation}
        \label{eq:watch-time}
        \bar \omega_t = \frac{\sum_{i=1}^{t}x_w[i]}{\sum_{i=1}^{t}x_v[i]}
    \end{equation}
    \item \textit{average watch percentage}$\bar\mu_t$: the average watch time $\bar\omega_t$ normalized by video duration $D$
    \begin{equation}
        \label{eq:watch-percentage}
        \bar\mu_t = \frac{\bar\omega_t}{D}
    \end{equation}
\end{itemize}

$\bar\omega_t$ is a positive number bounded by the video length, whereas $\bar\mu_{t}$ takes values between 0 and 1 and represents the average percentage of video watched.

\noindent\textbf{Engagement map.}
We observe that video duration is an important covariate on watch percentage.
In the \textsc{Tweeted videos} dataset, duration alone explains more than $58\%$ of the variance of watch percentage.
Intuitively, longer videos are less likely to be fully watched compared to shorter videos due to the limited human attention span.

\begin{figure}[t]
    \centering
    \subfloat{\includegraphics[width=0.235\textwidth]{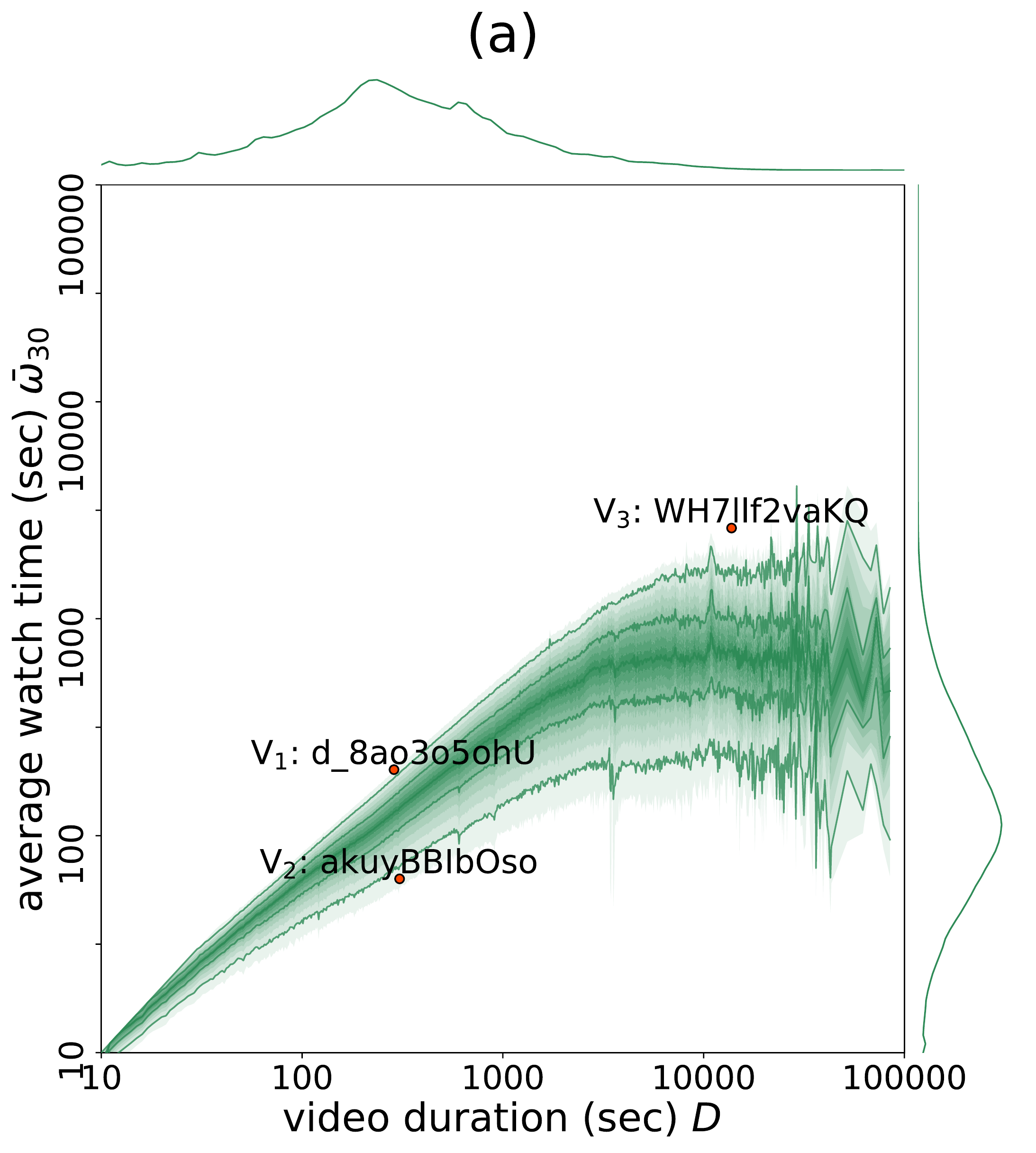}}
    \subfloat{\includegraphics[width=0.235\textwidth]{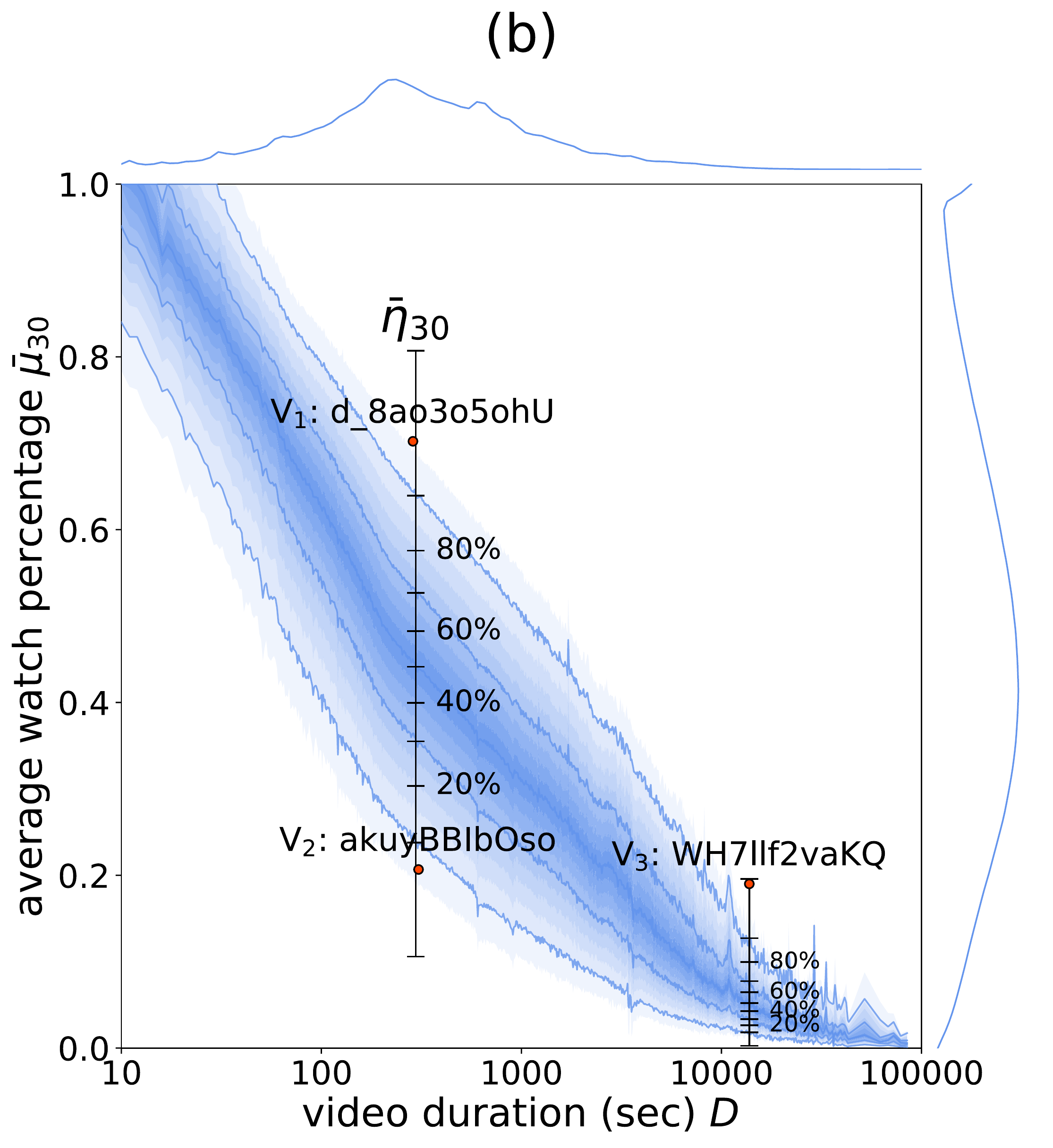}}
    \caption{Video engagement in the \textsc{Tweeted videos} dataset at the age of 30 days. (a) video duration $D$ \textit{vs} average watch time $\bar\omega_{30}$; (b) the \textit{engagement map}: video duration $D$ \textit{vs} average watch percentage $\bar \eta_{30}$.}
    \label{fig:eng-wt-map}
\end{figure}

We construct two 2-dimensional maps, where the x-axis shows video duration $D$, and the y-axis shows average watch time $\bar\omega_{30}$ (Fig.~\ref{fig:eng-wt-map}a) and average watch percentage $\bar\mu_{30}$ (Fig.~\ref{fig:eng-wt-map}b) over the first 30 days.
We project all videos in the \textsc{Tweeted videos} dataset onto both maps.
The x-axis is split into 1,000 equally wide bins in log scale.
We choose 1,000 bins to trade-off enough data in each bin and having enough bins.
We have also tried discretizing to smaller or larger number of bins, and the results are visually similar.
We merge bins containing a very low number of videos ($<$50) to nearby bins.
Overall, each bin contains between 50 and 38,508 videos.
The color shades correspond to data percentiles inside each bin: the darkest color corresponds to the median value and the lightest correspond to the extremes (0$\%$ and 100$\%$).
Both maps calibrate watch time and watch percentage against video durations: highly-watched videos are positioned towards the top of allocated bin, while barely-watched videos are at the bottom compared to other videos with similar length.

Those two maps are logically identical because the position of each video in Fig.~\ref{fig:eng-wt-map}b can be obtained by normalizing with its duration in Fig.~\ref{fig:eng-wt-map}a.
It is worth noticing that a linear trend exists between average watch time and video duration in the log-log space, with an increasing variance as duration grows.
In this work, we predominantly use the map of watch percentage (Fig.~\ref{fig:eng-wt-map}b) given its y-axis is bounded between [0,1], making it easier to interpret.
We denote this map as the \textit{engagement map}.

Note that our method of constructing the engagement map resembles the idea of non-parametric quantile regression, which essentially computes a quantile regression fit in an equally spaced span~\cite{koenker2005quantile}.
For smaller datasets, using quantile regression may result in a smoother mapping.
We tried quantile regression on \textsc{Tweeted videos} dataset, and we found that the values on both tails are inaccurate as the polynomial fits do not accurately reflect nonlinear trends. 
Our binning method works better in this case.
Finally, we remarks that the engagement map can be constructed at different ages, which allows us to study the temporal evolution of engagement (Sec.~\ref{ssec:temp-dynamics}).

\noindent\textbf{Relative engagement $\bar\eta_{t}$.}
Based on the engagement map, we propose the relative engagement $\bar\eta_{t} \in [0,1]$, defined as the rank percentile of video in its duration bin.
This is an average engagement measure in the first $t$ days.
Fig.~\ref{fig:eng-wt-map}b illustrates the relation between video duration $D$, watch percentage $\bar\mu_{30}$ and relative engagement $\bar\eta_{30}$ for three example videos.
Video $v_1$ ({\texttt{d\_8ao3o5ohU}}) shows kids doing karate and $v_2$ ({\texttt{akuyBBIbOso}}) is about teaching toddlers colors.
They are both about 5 minutes, but have different watch percentages, $\bar\mu_{30}(v_1)$= 0.70 and $\bar\mu_{30}(v_2)$=0.21.
These amount to very different values of the relative engagement: $\bar\eta_{30}(v_1)$=0.96, while $\bar\eta_{30}(v_2)$=0.07.
Video $v_3$ ({\texttt{WH7llf2vaKQ}}) is a much longer video ($D$=3 hours 49 minutes) showing a live fighting show.
It has a relatively low watch percentage ($\bar\mu_{30}(v_3)$=0.19), similar to $v_2$.
However, its relative engagement $\bar\eta_{30}(v_3)$ amounts to 0.99, positioning it among the most engaging videos in its peer group.

We denote the mapping from watch percentage $\bar\mu_{t}$ to relative engagement $\bar\eta_{t}$ as $f$, and its inverse mapping as $f^{-1}$.
Here $f$ is implemented as a length-1,000 look up table with a maximum resolution of 0.1\% (or 1,000 ranking bins).
For a given video with duration $D$, we first map it to corresponding bin on the engagement map, then return the engagement percentile by watch percentage.
Eq.~\ref{eq:engagement-interchange} describes the mapping between relative engagement and average watch percentage using engagement map.

\begin{equation}
    \label{eq:engagement-interchange}
    \bar\eta_t = f(\bar\mu_t, D) \Leftrightarrow \bar\mu_t = f^{-1}(\bar\eta_t, D)
\end{equation}

While researchers have observed that watch percentage is affected by video duration~\cite{guo2014video,park2016data}, to the best of our knowledge, this work is the first to quantitatively map its non-linear relation with video duration and present measurements in a large-scale dataset.

\subsection{Relative engagement and video quality}
\label{ssec:video-quality}

\begin{figure}[t]
    \centering
    \subfloat{\includegraphics[width=0.235\textwidth]{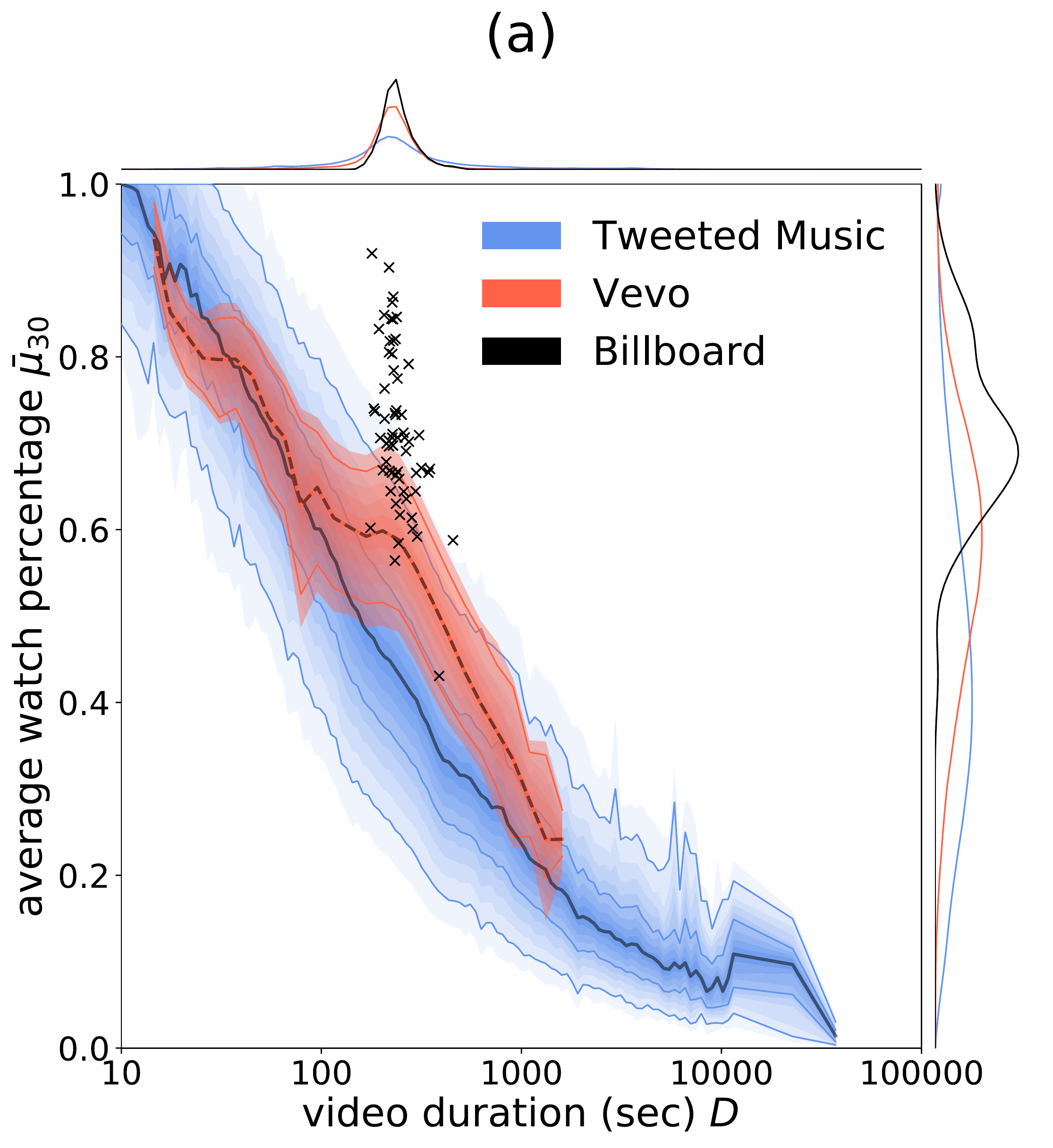}}
    \subfloat{\includegraphics[width=0.235\textwidth]{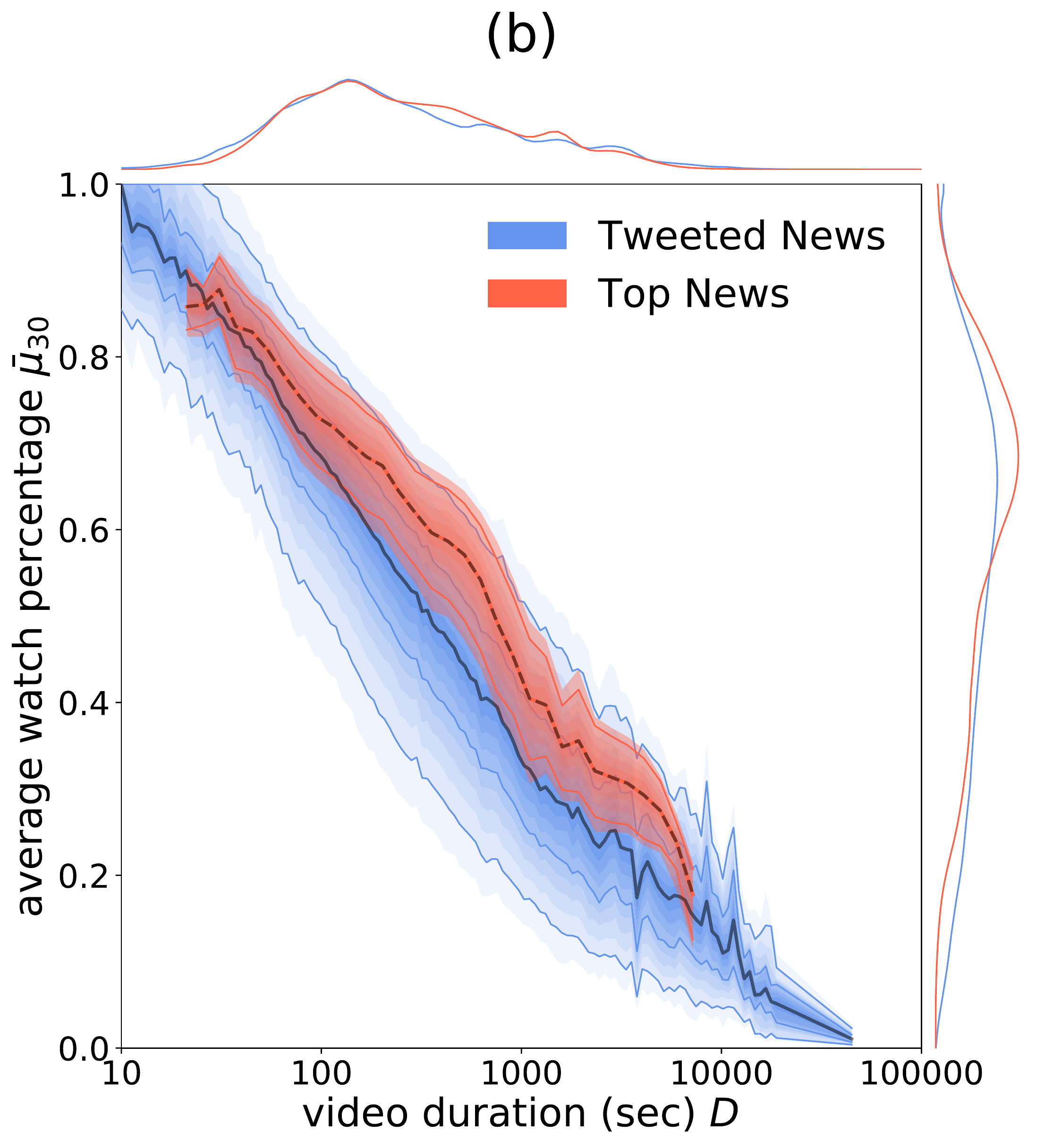}}
    \caption{Relative engagement and video quality for Music (a) and News (b). Videos in \textsc{Quality videos} dataset are shifted towards higher relative engagement compared to that in \textsc{Tweeted videos}. Best viewed in colors.}
    \label{fig:eng-quality}
\end{figure}

We examine the relation between relative engagement and video quality.
We place the \textsc{Quality videos} datasets (Sec.~\ref{ssec:datasets}) on the engagement map.
Fig.~\ref{fig:eng-quality}a plots the engagement map of all Music videos in the \textsc{Tweeted Videos} (blue), that of the \textsc{Vevo Videos} (red), and the videos in the \textsc{Billboard videos} as a scatter plot (black dots).
Similarly, Fig.~\ref{fig:eng-quality}b plots the engagement map of all News videos in the \textsc{Tweeted Videos} in blue and that of the \textsc{Top News Videos} in red.
All the maps are built from observations in the first 30 days.

Visibly, the \textsc{Quality Videos} are skewed towards higher relative engagement values in both figures. 
Most notably, 44 videos in the \textsc{Billboard Videos} dataset (70\% of the dataset) possess a high relative engagement of over 0.9.
The other 30\% of videos have an average $\bar\eta_{30}$ of 0.83 with a minimum of 0.54.
For \textsc{Quality videos}, the 1-dimensional density distribution of average watch percentage $\bar\mu_{30}$ also shifts to the upper end as shown on the right margin of Fig.~\ref{fig:eng-quality}.
Overall, relative engagement values are high for content judged to be high quality by experts and the community.
Thus, relative engagement is one plausible surrogate metric for content quality.

\begin{figure}[t]
    \centering
    \includegraphics[width=0.47\textwidth]{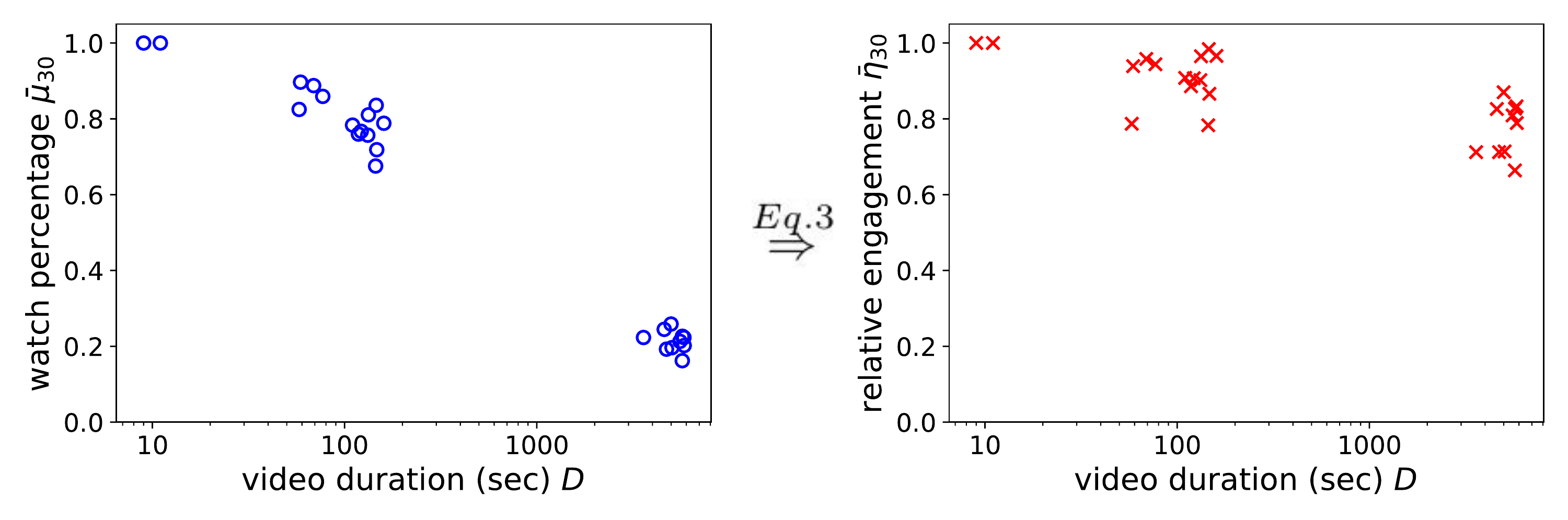}
    \caption{Watch percentage $\bar \mu_{30}$ (left) and relative engagement $\bar \eta_{30}$ (right) for videos in channel \texttt{PBABowling}. While it appears that $\bar \mu_{30}$ has a linear relation with the logarithmic duration $\log_{10} D$, $\bar \eta_{30}$ can be reasonably explained by only using the mean value of $\bar \eta_{30}$.}
    \label{fig:eng-channel}
\end{figure}

\noindent\textbf{Relative engagement within channel.}
Fig.~\ref{fig:eng-channel} shows the engagement mapping results of 25 videos within one channel (\texttt{PBABowling}).
This channel uploads sports videos about \textit{Professional Bowlers Association} with widely varying lengths -- from 2-minute player highlights to 1-hour event broadcasts.
Video length has a significant impact: the short video cluster has mean average watch percentage $\bar\mu_{30}$ of 0.82, whereas the long video cluster has mean $\bar\mu_{30}$ of 0.21.
However, after mapping to relative engagement, those two clusters have mean $\bar\eta_{30}$ of 0.92 and 0.78 -- much more consistent for this channel than measured by watch percentage.
Overall, the mean relative engagement of channel \texttt{PBABowling} is 0.86, which suggests this channel is likely to produce more engaging videos than an average YouTube channel, regardless of the video length.
This example illustrates video relative engagement tends to be stable within the same channel, and sheds some light on using past videos to predict future relative engagement.

\subsection{Temporal dynamics of relative engagement}
\label{ssec:temp-dynamics}

\textit{How does engagement change over time?}
This question is important because popularity dynamics tend to be bursty and hard to predict~\cite{cheng2014can}.
If engagement dynamics can be shown to be stable, it is useful for content producers to understand watch patterns from early observation.
Note that the method for constructing the engagement map is the same, but one can use data at different ages $t$ to build different mapping function $f(\bar\mu_t, D)$.

\begin{figure}[t]
    \centering
    \includegraphics[width=0.47\textwidth]{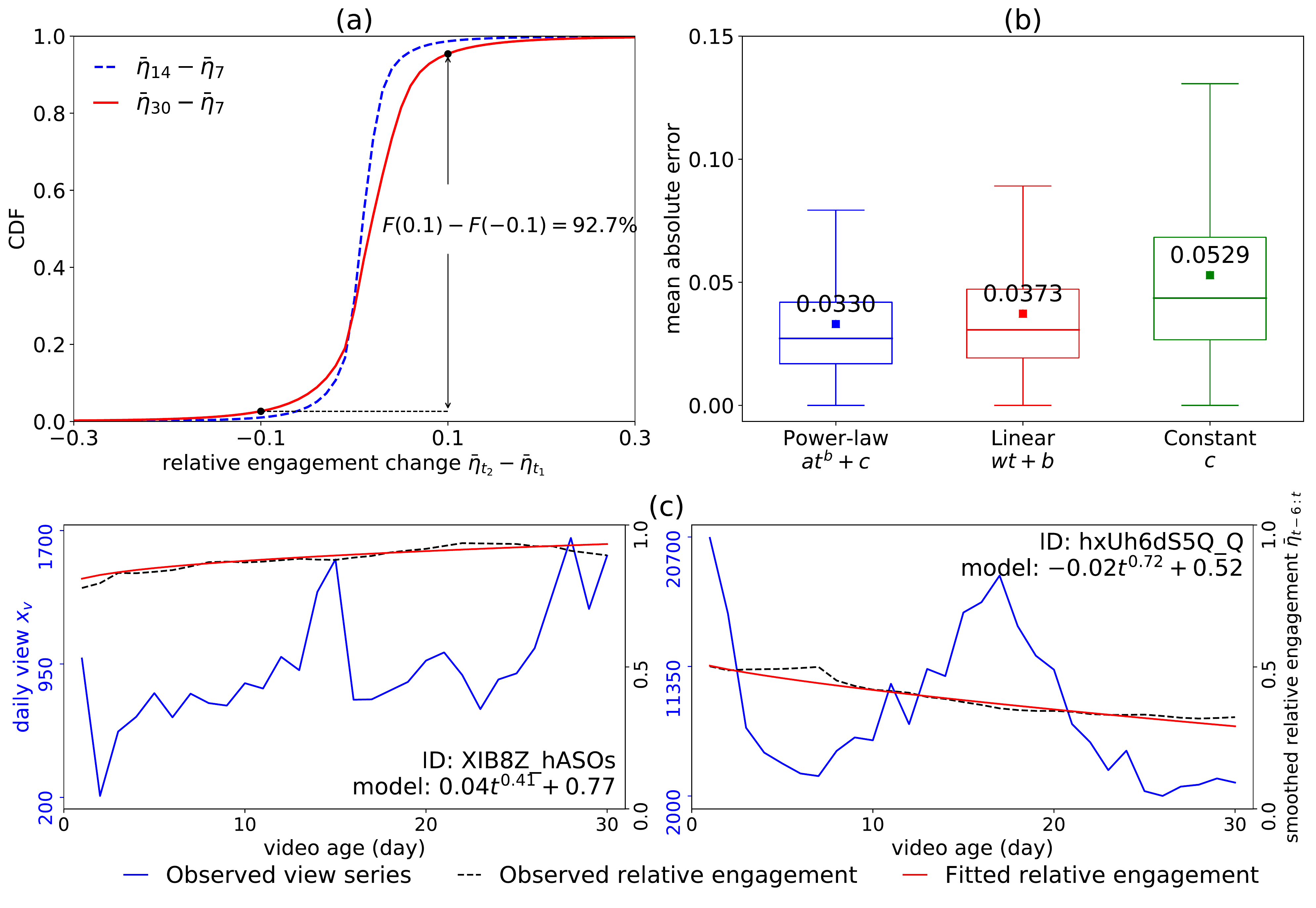}    
    \caption{Relative engagement is stable over time. (a) CDF of temporal change in relative engagement of day 7 \textit{vs} day 14 (blue dashed), day 7 \textit{vs} day 30 (red solid). (b) Fitting error of power-law model (blue), linear regressor (red) and constant function (green) in \textsc{Tweeted videos}. (c) Temporal view series (blue) and smoothed daily relative engagement (black dashed) fitted by generalized power-law model $a t^b + c$ (red).}
    \label{fig:temporal-fitting}
\end{figure}

\noindent\textbf{Relative engagement is stable over time.}
We examine the temporal change of relative engagement at two given days $t_1$ and $t_2$ ($t_1$$<$$t_2$) in \textsc{Tweeted videos}.
We denote the cumulative distribution function (CDF) as $F_x(\Delta \bar \eta)$, where $x$$=$$\bar \eta_{t_2}$$-$$\bar \eta_{t_1}$.
This computes the fraction of videos with relative engagement changing \textit{less} than $\Delta \bar \eta$ during $t_1$ to ${t_2}$.
Fig.~\ref{fig:temporal-fitting}a shows $\Delta \bar \eta$ distribution of day 7 \textit{vs} day 14 and day 7 \textit{vs} day 30.
There are 4.6\% of videos that increase more than 0.1 and 2.7\% that decrease more than 0.1, yielding 92.7\% of the videos with an absolute relative engagement change of less than 0.1 between day 7 and day 30.
Such a small change results from the fact that relative engagement $\bar\eta_{t}$ is defined as average measure over the past $t$ days.
It suggests that future relative engagement can be predicted from early watch patterns within a small margin of error.
Similarly, this observation extends to both average watch percentage $\bar \mu_t$ and average watch time $\bar \omega_t$.

Next we examine relative engagement on a daily basis.
To avoid days with zero views, we use a 7-day sliding window, i.e., changing the summations in Eq.~\ref{eq:watch-time} to between $t$-6 and $t$, yielding a \textit{smoothed} daily watch percentage $\bar\mu_{t-6: t} = \frac{\sum_{i=t-6}^{t}x_w[i]}{D \sum_{i=t-6}^{t}x_v[i]}$.
We then convert $\bar\mu_{t-6: t}$ to smoothed daily relative engagement $\bar\eta_{t-6: t}$ via the corresponding engagement map.
For $t$$<$7, we calculate relative engagement from all prior days before $t$.

Fig.~\ref{fig:temporal-fitting}c shows the daily views and smoothed relative engagement over the first 30 days of two example videos.
While the view series has multiple spikes (blue), relative engagement is stable with only a slightly positive trend for video \texttt{XIB8Z\_hASOs} and a slightly negative trend for \texttt{hxUh6dS5Q\_Q} (black dashed).
View dynamics have been shown to be affected by external sharing behavior~\cite{rizoiu2017online}, the stability of relative engagement can be explained by the fact that it measures the average watch pattern but not how many people view the video.

\noindent\textbf{Fitting relative engagement dynamics.}
We examine the stability of engagement metrics across the entire \textsc{Tweeted Videos} dataset.
If the engagement dynamics can be modeled by a parametric function, one can forecast future engagement from initial observations.
To explore which function best describes the gradual change of relative engagement $\bar\eta_{t}$,
we examine generalized power-law model ($a t^b + c$)~\cite{yu2015lifecyle}, linear regressor ($w t + b$), and constant ($c$) function.
For videos in \textsc{Tweeted videos}, we fit each of the three functions to smoothed daily relative engagement series $\bar\eta_{t-6: t}$ over the first 30 days.
Fig.~\ref{fig:temporal-fitting}b shows that power-law function fits best on the dynamics of relative engagement, with an average mean absolute error of $0.033$.

To sum up, we observe that relative engagement $\bar\eta_t$ is stable throughout lifetime, which implies that early watch pattern is a strong predictor for future engagement.
Therefore, in the next section, we set up a prediction task to examine whether engagement can be predicted \textit{before} upload.

\section{Predicting engagement}
\label{sec:predict}

In this section, we predict relative engagement and watch percentage of a video \textit{before} it is uploaded.
We further analyze the relation between video features and engagement metrics.

\subsection{Prediction tasks setup}
\label{ssec:predtask}

We observe that relative engagement and watch percentage are stable over time(Sec.~\ref{ssec:temp-dynamics}), which makes them attractive prediction targets. 
Furthermore, it is desirable to predict them before videos get uploaded, and viewing or commenting behavior is observed. 

\noindent{\textbf{Prediction targets.}}
We setup two regression tasks to predict average watch percentage $\bar \mu_{30}$ and relative engagement $\bar \eta_{30}$.
Watch percentage is intuitively useful for content producers, while relative engagement is designed to calibrate watch percentage against duration as detailed in Sec.~\ref{ssec:engagement-score}.
It is interesting to see whether such calibration changes prediction performance.
We report three evaluation results: predicting relative engagement and watch percentage directly, and predicting relative engagement then mapping to watch percentage via engagement map by using Eq.~\ref{eq:engagement-interchange}.
We do not predict average watch time because it can be deterministically computed by multiplying watch percentage and duration.

\noindent{\textbf{Training and test data.}}
We split \textsc{Tweeted videos} at 5:1 ratio over publish time.
We use the first 51 days (2016-07-01 to 2016-08-20) for training, containing 4,455,339 videos from 1,132,933 channels;
and the last 11 days for testing (2016-08-21 to 2016-08-31), containing 875,865 videos from 366,311 channels.
242,017, or 66\% of channels in the test set have appeared in training set, however, none of the videos in the test set is in the training set.
The engagement map between watch percentage and relative engagement is built on the training set over the first 30 days.
We split the dataset in time to ensure that learning is on past videos and prediction is on future videos.

\noindent\textbf{Evaluation metrics.}
Performance is measured with two metrics:
\begin{itemize}
    \item Mean Absolute Error
    $MAE = \frac{1}{N}\sum_{i=1}^{N}|y_i - \hat{y_i}|$
    \item Coefficient of Determination 
    $ R^2 = 1 - \frac{\sum_{i=1}^{N}(y_i - \hat y_i)^2}{\sum_{i=1}^{N}(y_i - \bar y)^2}$
\end{itemize}
Here $y$ is the true value, $\hat y$ the predicted value, $\bar y$ the average; $i$ indexes samples in the test set.
MAE is a standard metric for average error.
$R^2$ quantifies the proportion of the variance in the dependent variable that is predictable from the independent variable~\cite{allen1997coefficient}, and is often used to compare different prediction problems~\cite{martin2016exploring}.
A lower MAE is better whereas a higher $R^2$ is better.

\subsection{Features}
\label{ssec:predfeat}

We describe each YouTube video with 4 types of features as summarized in Table~\ref{table:features}.

\noindent\textbf{Control variable.}
Because video duration is the primary source of variation for engagement (Fig.~\ref{fig:eng-wt-map}), we use \textit{duration} as a control variable and include it in all predictors.
In \textsc{Tweeted Videos} dataset, durations vary from 1 second to 24 hours, with a mean value of 12 minutes and median of 5 minutes.
We take the logarithm (base 10) of duration to account for the skew.

\noindent\textbf{Context features} are provided by video uploader.
They describe basic video properties and production quality \cite{hessel2017cats}.
\begin{itemize}
    \item \textit{Definition:}``1" represents high definition (720p or 1080p) and "0" represents low definition (480p, 360p, 240p or 144p).
    High definition yields better perceptual quality and encourages engagement~\cite{dobrian2011understanding}. 
    \item \textit{Category:} broad content identifications assigned by video producers, the full list is shown in Table~\ref{table:data} (bottom).
    Here we encode it as an 18-dimensional one-hot vector.
    \item \textit{Language:} we run \texttt{langdetect} package on the video description and choose the most likely language.
    \texttt{langdetect} implements a Naive Bayes classifier to detect 55 languages with high precision~\cite{nakatani2010langdetect}.
    The language is indicative of audience demographics.
\end{itemize}

\begin{table}
    \centering
    \begin{tabular}{lp{5cm}}
        \toprule
        \multicolumn{2}{c}{\textbf{Control variable (D)}}\\
        \textit{Duration} & Logarithm of duration in seconds\\
        \midrule
        \multicolumn{2}{c}{\textbf{Context features (C)}}\\
        \textit{Definition} & Binary, high definition or not\\
        \textit{Category} & One hot encoding of 18 categories\\
        \textit{Language} & One hot encoding of 55 languages\\
        \midrule
        \multicolumn{2}{c}{\textbf{Freebase topic features (T)}}\\
        \textit{Freebase topics} & One hot sparse representation of 405K topics\\
        \midrule
        \multicolumn{2}{c}{\textbf{Channel reputation features (R)}}\\
        \textit{Activity level} & Mean number of daily upload\\
        \textit{Past engagement} & Mean, std and five points summary of previously uploaded videos\\
        \midrule
        \multicolumn{2}{c}{\textbf{Channel specific predictor (CSP)}}\\
        \multicolumn{2}{c}{One predictor for each channel using available features}\\
        \bottomrule 
    \end{tabular}
    \caption{Overview of features for predicting engagement.}
    \label{table:features}
\end{table}

\noindent\textbf{Freebase topics features.}
YouTube labels videos with Freebase entities~\cite{bollacker2008freebase}. 
These labels incorporate user engagement signals, video metadata and content analysis~\cite{youtube8m}, and are built upon a large amount of data and computational resources.
With the recent advances in computer vision and natural language processing, there may exist more accurate methods for annotating videos.
However, one can not easily build such an annotator at scale, and finding the best video annotation technique is beyond the scope of this work.
On average, each video in the \textsc{Tweeted Videos} dataset has 6.16 topics. 
Overall, there are 405K topics and 98K of them appear more than 6 times.
These topics vary from broad categories (\texttt{Song}), to specific object (\texttt{Game of Thrones}), celebrities (\texttt{Adele}), real-world events (\texttt{2012 Seattle International Film Festival}) and many more.
Such fine-grained topics are descriptive of video content.
While learning embedding vectors can help predict engagement~\cite{covington2016deep},
using raw Freebase topics enables us to interpret the effect of individual topic (Sec.~\ref{ssec:topicana}).

\noindent\textbf{Channel reputation features.}
Prior research shows that user features are predictive for product popularity~\cite{martin2016exploring,mishra2016feature}.
Here we compute feature from a channel's history to represent its reputation.
We could not use social status indicators such as the number of subscribers, because it is a time-varying quantity and the value when a video is uploaded can not be retrospectively obtained.
Thus, we compute two proxies for describing channel features.
\begin{itemize}
    \item \textit{Activity level:} mean number of daily published videos by channels in the training data.
    Intuitively, channels with higher upload rates reflect better productivity.
    \item \textit{Past engagement:} relative engagement of previously uploaded videos from the same channel in the training set.
    Here we compute mean, standard deviation and five points summary: median, 25th and 75th percentile, min and max. 
\end{itemize}

Several features used in prior works are interesting, but they do not apply in our setting.
Network traffic measurement~\cite{dobrian2011understanding} requires access to the hosting backend. 
Audience reactions such as likes and comments~\cite{park2016data} can not be obtained before a video's upload.

\subsection{Prediction methods and results}
\label{ssec:predresult}

\noindent\textbf{Prediction methods.}
We use linear regression with L2-regularization to predict engagement metrics, $\bar\eta_{30}$ and $\bar\mu_{30}$, both lie between 0 and 1. 
Since the dimensionality of Freebase topics features is high (4M x 405K), we convert the feature matrix to a sparse representation, allowing the predictor to be trained on one workstation.
We adopt a fall-back strategy to deal with missing features.
For instance, we use the context predictor for videos for which the channel reputation features are unavailable.
The fall-back setting usually results in a lower prediction performance, however it allows to predict engagement for \textit{any} video.
We also tried KNN regression and support vector regression, but they did not yield better performances.

\begin{figure}
    \centering
    \includegraphics[width=0.47\textwidth]{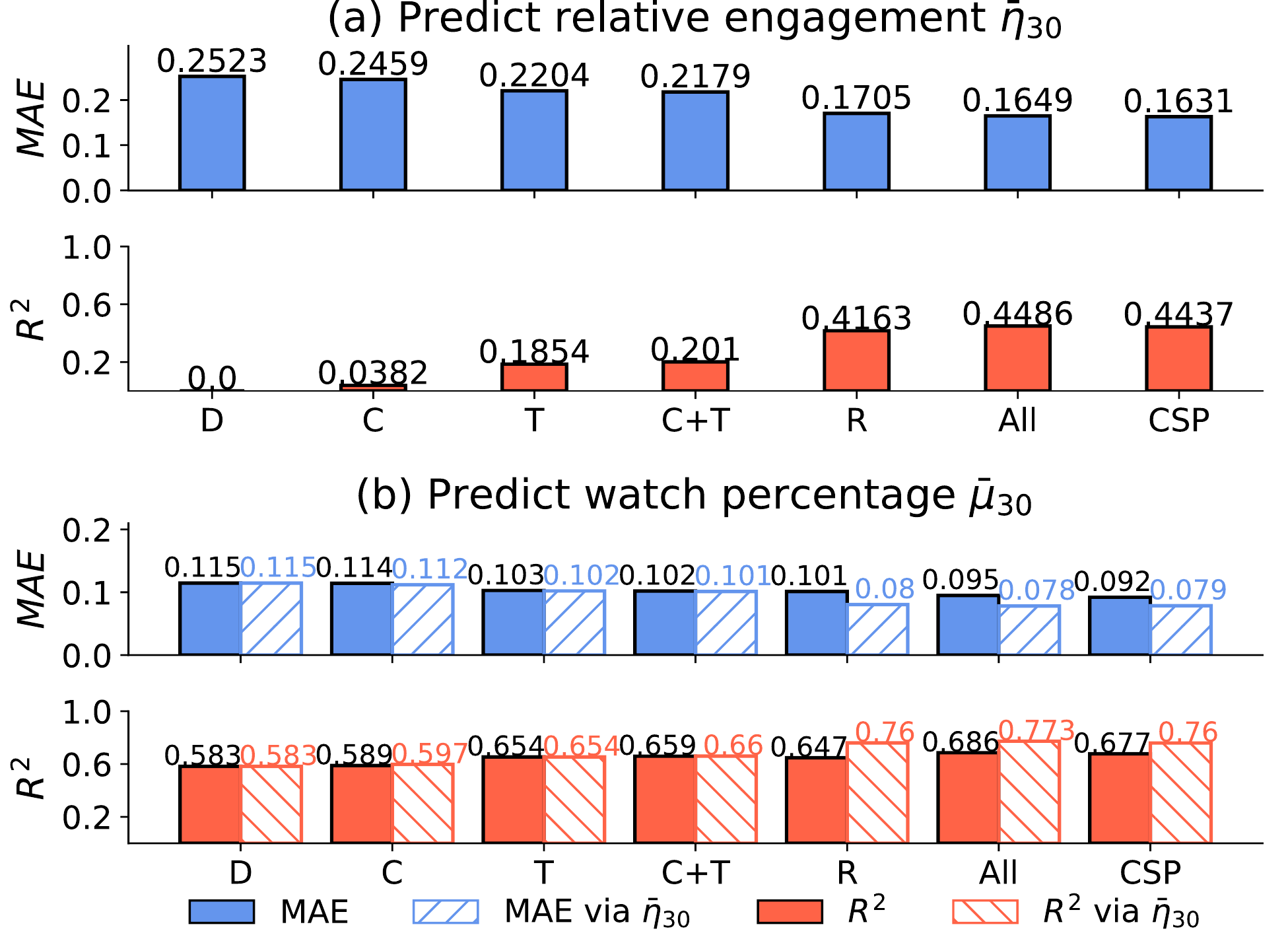}
    \caption{Summary of engagement prediction with two metrics, $MAE$: lower is better; $R^2$: higher is better. (a): Performance for predicting $\bar\eta_{30}$ in different feature combinations. (b): Performance for predicting $\bar\mu_{30}$ in different feature combinations, directly (solid bars, left) or via relative engagement $\bar\eta_{30}$ (shaded bars, right). Predicting watch percentage via converting relative engagement performs better than predicting watch percentage directly in all predictors.}
    \label{fig:prediction-results}
\end{figure}

\noindent\textbf{Channel specific predictor (CSP).}
In addition to the shared predictor, we train a separate predictor for each channel that has at least 5 videos in the training set.
This fine-grained predictor covers 61.4\% videos in the test data and may capture the ``on-topic'' effect within channel~\cite{martin2016exploring}.
Intuitively, a channel might have specialty on certain topics and videos about those attract the audience to watch longer.
For the remaining 38.6\% videos, we use the shared linear regressor with all available features.

\noindent\textbf{Prediction results.}
Fig.~\ref{fig:prediction-results}a summarizes the results of predicting the relative engagement $\bar\eta_{30}$.
Context (\textbf{C}) and Freebase topics (\textbf{T}) alone are weak predictors, explaining 0.04 and 0.19 variance of $\bar\eta_{30}$ in the test set.
Combining the two (\textbf{C+T}) yields a slight gain over Freebase topics.
Channel reputation (\textbf{R}) is the strongest feature, achieving $R^2$=0.42, and is slightly improved by adding context and Freebase topics.
Channel-specific predictor (\textbf{CSP}) performs similarly to the All-feature predictor (\textbf{All}), suggesting that one can use a shared predictor to achieve similar performance with finer-grained per-channel model for this task.

Average watch percentage $\bar\mu_{30}$ is easier to predict, achieving $R^2$ up to 0.69 (Fig.~\ref{fig:prediction-results}b) by using all features.
Interestingly, predicting $\bar\eta_{30}$ then mapping to $\bar\mu_{30}$ consistently outperforms direct prediction of $\bar\mu_{30}$, achieving $R^2$ of 0.77.
This shows that removing the influence of video duration via engagement map is beneficial for predicting engagement.

\begin{figure}
    \centering
    \includegraphics[width=0.47\textwidth]{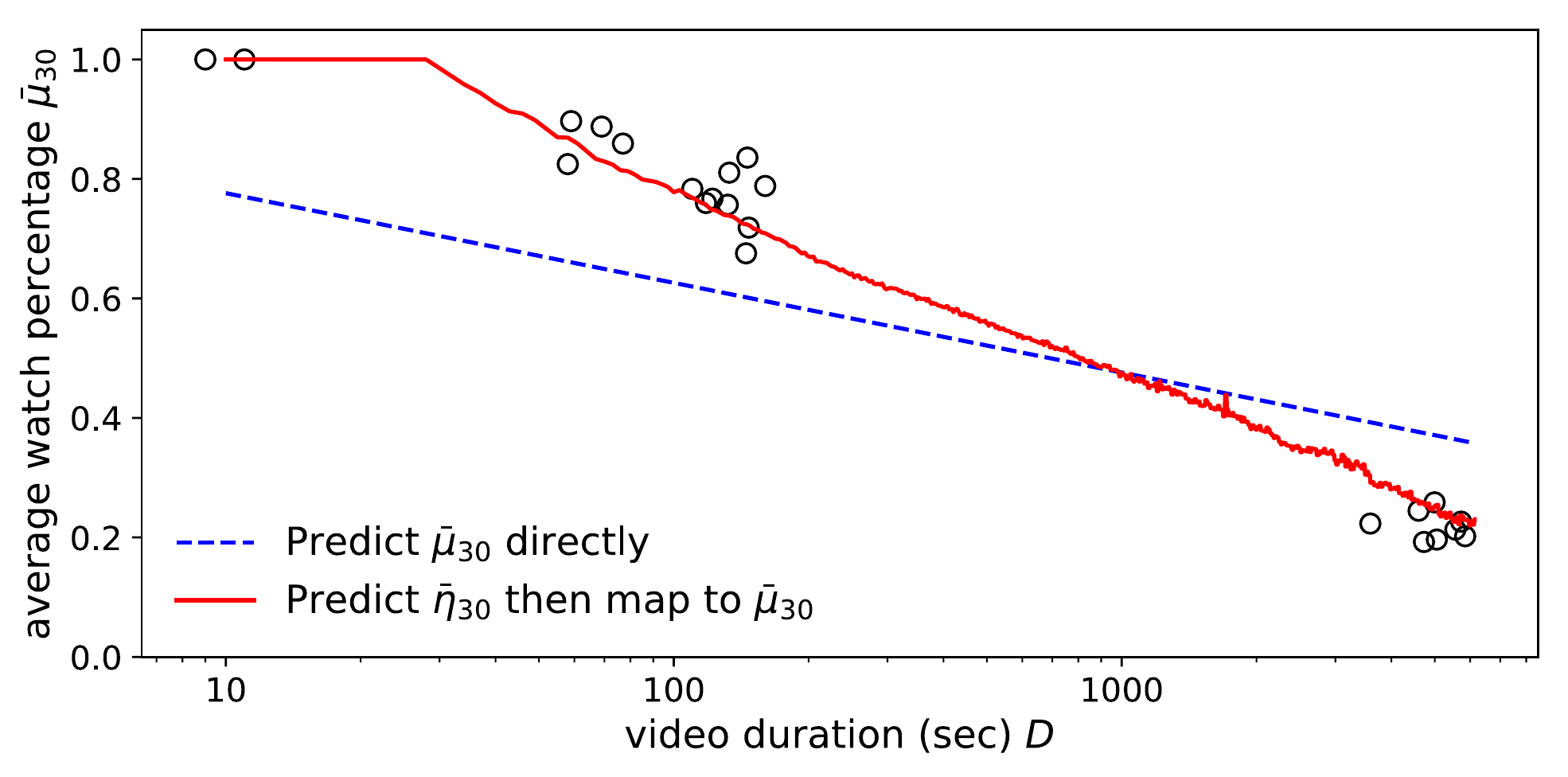}
    \caption{Shared linear regressors with channel reputation features on channel \texttt{PBABowling}, for predicting $\bar\mu_{30}$ (blue dashed) and predicting $\bar\eta_{30}$ then mapping to $\bar\mu_{30}$ (red solid).}
    \label{fig:prediction-explain}
\end{figure}

To understand why predicting via $\bar\eta_{30}$ performs better, we examine the shared linear regressors in both tasks.
For simplicity, we include video duration and channel reputation features as covariates, and exclude the (generally much weaker) context and Freebase topics features for this example.
In Fig.~\ref{fig:prediction-explain}, we visualize the two shared channel reputation predictors (\textbf{R}) at different video lengths for channel \texttt{PBABowling} (also shown in Fig.~\ref{fig:eng-channel}): one predicts $\bar\mu_{30}$ directly (blue dashed), and the other predicts $\bar\eta_{30}$, then maps to $\bar\mu_{30}$ via the engagement map (red solid).
The engagement map captures the non-linear effect for both short and long videos.
In contrast, predicting $\bar\mu_{30}$ directly does not capture the bimodal duration distribution here: it overestimates for longer videos and underestimates for shorter videos.

\noindent\textbf{Analysis of failed cases.}
We investigate the causes of failed prediction for each predictor.
The availability of channel information appears important -- for most poorly predicted videos, their channels have only one or two videos in the training set.
Moreover, some topics appear more difficult to predict than others.
For example, videos that are labeled with \texttt{music} obtain a \textit{MAE} score of 0.175 ($\bar\eta_{30}$ using the All-feature predictor).
This amounts to an error increase of 28\% compared to videos labeled with \texttt{obama} (\textit{MAE} = 0.136). 
Lastly, the prediction performance varies considerably even for videos from the same channel and identically labeled.
For example, the channel \texttt{Smyth Radio} (\texttt{UC79quCUqSgHyAY9Kwt1V6mg}) released a series of videos about ``United States presidential election'', 8 of which are in our dataset: 6 are in the training set and 2 are in the test set.
These videos have similar lengths (3 hours) and they are produced in a similar style. 
The 6 videos in training set are watched on average between 3 and 10 minutes, yielding a $\bar\eta_{30}$ of 0.08.
However, the 2 videos in the test set achieve considerable attention -- 1.5 hours watch time on average, projecting $\bar\eta_{30}$ at 1.0.
One possible explanation is that the videos in the test set discuss conspiracy theories and explicitly lists them in the title.

Overall, engagement metrics are predictable from context, topics and channel information in a \textit{cold-start} experiment setting.
Although channel reputation information is the strongest predictor, Freebase topics features are also somewhat predictive.

\subsection{Are Freebase topics informative?}
\label{ssec:topicana}

In this section, we analyze the Freebase topics features in detail and provide actionable insights for producing videos.
Firstly, we group videos by Freebase topic and extract the most frequent 500 topics.
Next we measure the amount of information gain with respect to relative engagement conditional entropy, defined in following equation:

\begin{equation}
    \label{eq:ce}
    \scalebox{0.93}[1]{$H(Y|X_i=1) = -\sum_{y \in Y} P(y|x_i=1) \log_2 P(y|x_i=1)$}
\end{equation}

\begin{figure}
    \centering
    \includegraphics[width=0.47\textwidth]{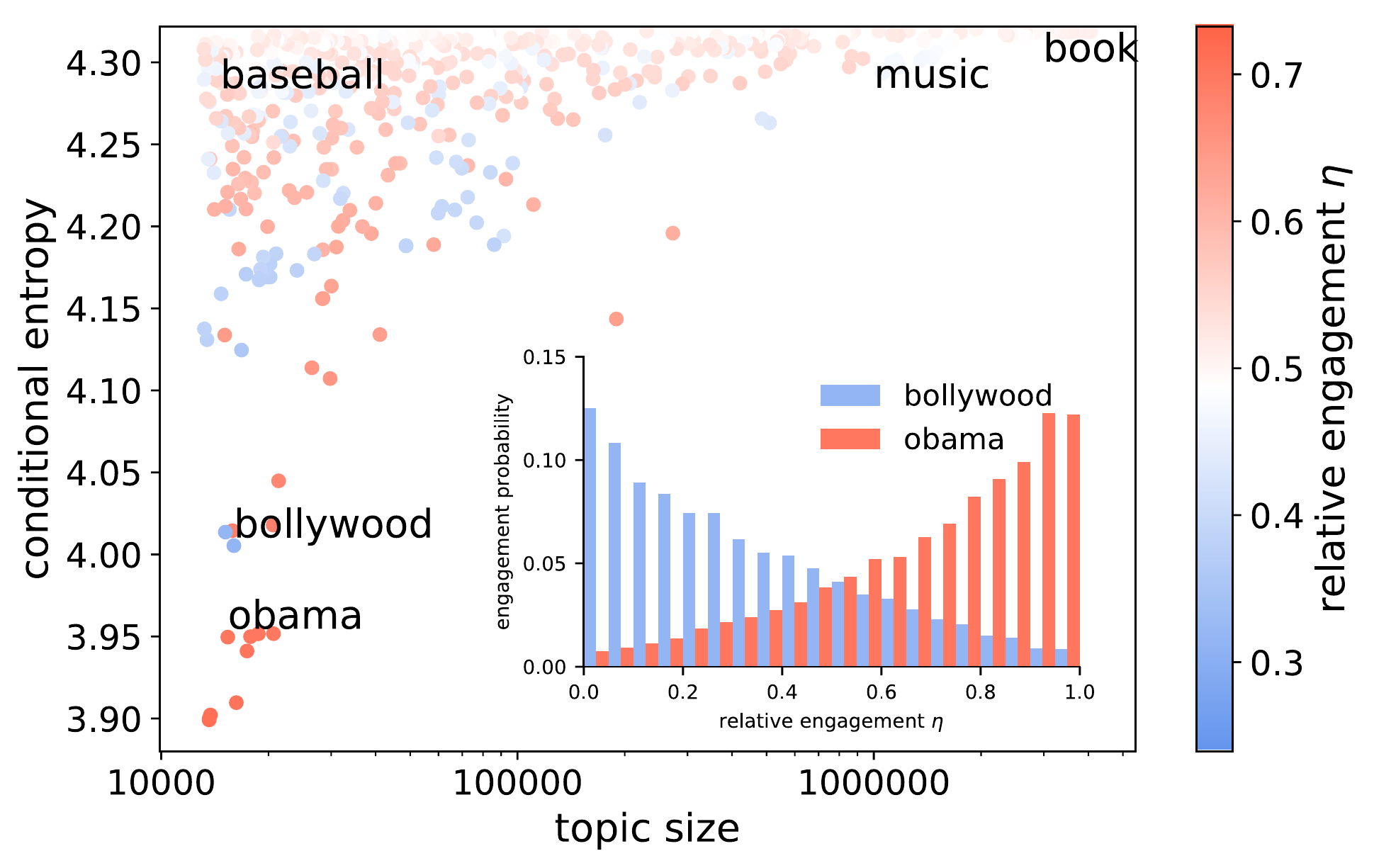}
    \caption{Informativeness for the most frequent 500 Freebase topics, measured by conditional entropy. (Inset) relative engagement distribution for two example topics: \texttt{Obama} - high engagement; \texttt{bollywood} - low engagement.}
    \label{fig:topics}
\end{figure}

Each topic is represented as a binary variable $x_i \in \{0, 1\}$, for $i=1,\ldots,500$.
We divide relative engagement into 20 bins, and $y$ is the discretized bin.
A lower conditional entropy indicates the presence of current topic is informative for engagement prediction (either higher or lower).
Here we calculate $H(Y|X=1)$ rather than $H(Y|X)$, because $X$=0 represents the majority of videos for most topics and the corresponding term will dominate.
Using $H(Y|X=1)$ quantifies its effect only when the topic is in presence~\cite{sedhain2013social}.
Fig.~\ref{fig:topics} is a scatter plot of topic size and conditional entropy.
Here large topics such as \texttt{book} (3.2M videos) or \texttt{music} (842K videos) have high conditional entropy and mean relative engagement close to 0.5,
which suggests they are not informative in predicting engagement.
All informative topics (e.g., with conditional entropy 4.0 and lower) are relative small (e.g., appearing around 10K times in the training set).
Fig.~\ref{fig:topics} (inset) plots two example topics that are very informative on engagement, from which we observe that videos about \texttt{bollywood} are more likely to have a low relative engagement while topic \texttt{obama} tends to keep audience watching longer.
However, not all small topics are informative.
A counter-example is \texttt{baseball}, which has a small topic size but a high condition entropy value.

In summary, watch percentage and relative engagement are predictable in a \textit{cold-start} setting, before any behavioral data is collected.
A few content-based semantic topics are predictive of low- or high- engagement.
Such observation can help content producers make more engaging videos.

\section{Related work}

\noindent\textbf{Measuring engagement in online content.}
Many researchers have analyzed engagement behavior towards web content.
For example, the line of work that measures web page reading pattern often exploits auxiliary toolkit such as mouse-tracking~\cite{arapakis2014understanding} instrumented browsers.
In search engine and recommender systems, dwell time, which is conceptually close to video watch time, has been widely used~\cite{covington2016deep}.
Interestingly, \cite{yi2014beyond} compared two systems that optimize for clicks and dwell time, and found the one towards dwell time achieved better performance on ranking relevant products.
All the above works focus on engagement with an individual user.
However, user-level data is unavailable to content producers on YouTube platform.
Our work measures engagement at an aggregate level, as complementary to individual engagement study.

The work most relevant to ours on measuring video aggregate engagement is from~\cite{park2016data}, in which the authors show the predictive power of collective reactions (e.g., view, like, and comment sentiment) for predicting average watch percentage.
However, these features require observing videos for some period of time.
Most importantly, a large fraction of videos do not have comments ~\cite{cheng2008statistics}, making this prediction setup inapplicable to a random YouTube video.
In contrast, our work is the first to quantitatively measure the effect of video duration over a large-scale dataset and predict watch percentage in a \textit{cold-start} setup.
We further discuss related works in the following three directions.

\noindent\textbf{Estimating quality of online content.}
\texttt{MusicLab experiment} is the first to measure online content quality in an experimental environment~\cite{salganik2006experimental}, in which they measure as the fraction of download number over listening number.
This experiment is further studied by~\cite{krumme2012quantifying}, who propose a two-step process to characterize user behavior in social systems.
The key influencing factor in the first step is popularity such as product appeal and market position, while the second step is merely affected by content quality.
\cite{stoddard2015popularity} has measured this process in Reddit and Hacker News.
In this work, our notions of popularity and engagement are inspired by this two-step process, intuitively describing \textit{the decision to click} and \textit{the decision to interact} on YouTube.
Moreover, \cite{van2016aligning} show that popularity is a poor proxy to represent quality in online market.
Thus, we propose a new metric \textit{relative engagement} based on the engagement step, and formalize it to correlate with video quality.

\noindent\textbf{Explaining popularity towards online videos.}
One of the most studied attributes is video popularity dynamics, defined as the number of times they are viewed.
A number of models have been proposed to describe the popularity dynamics, such as a series of endogenous relaxations~\cite{crane2008robust} or multiple power-law phases~\cite{yu2015lifecyle}.
Other studies link popularity dynamics to epidemic contagion~\cite{bauckhage2015viral}, external stimulation~\cite{yu2014twitter} or geographic locality~\cite{Brodersen2012a}.
However, the amount of time that videos are watched has mainly been overlooked, despite becoming the centric metric for recommendation in YouTube~\cite{Meyerson} and Facebook~\cite{Bapna}.
In this work, we provide an in-depth study on video engagement dynamics, and investigate key influencing factors.

\section{Conclusion}

In this paper, we measure a set of aggregate engagement metrics for online videos, including average watch time, average watch percentage, and a new metric, \textit{relative engagement}.
We study the proposed metrics on a publicly available dataset of 5.3 million videos.
We show that relative engagement is stable over the video lifetime, and strongly correlates with established notions of video quality.
In addition, we show average watch percentage can be predicted (with $R^2$=0.77) from public information, such as video context, topics, and channel, without observing any user reaction.
This is a significant result that separates the tasks of estimating engagement with predicting popularity over time.

\noindent\textbf{Limitations.}
Our observations are only on publicly available videos. 
It is possible that untweeted, private and unlisted videos behave differently.
The attention data used are aggregated over all viewers of a video.
Therefore our observations are more limited than those from content hosting site that has individual user attributes and reactions.
Hence our results do not directly translate to user-specific engagement.

\noindent\textbf{Future work and broader implications.}
For future work, one open problem is to quantify the gap between aggregate and individual measurements.
Another is to extract more sophisticated features and to apply more advance techniques to improve the prediction performance.
The observations in this work provide content producers with a new set of tools to create engaging videos and forecast user behavior.
For video hosting sites, engagement metrics can be used to optimize recommender systems and advertising strategies, as well as to detect potential clickbaits.

\noindent\textbf{Acknowledgments.}
This research is sponsored by the Air Force Research Laboratory, under agreement number FA2386-15-1-4018.
We thank National eResearch Collaboration Tools and Resources (Nectar) for providing computational resources, supported by the Australian Government.

\fontsize{9.0pt}{10.0pt}
\selectfont
\bibliography{yt-engagement-ref}
\bibliographystyle{aaai}

\end{document}